\documentclass[10pt,english,preprint]{elsarticle}
\usepackage[T1]{fontenc}
\usepackage[latin9]{inputenc}
\usepackage{amsmath}
\usepackage{graphicx}
\usepackage{amssymb}
\usepackage{esint}

\makeatletter

\providecommand{\tabularnewline}{\\}

\@ifundefined{definecolor}
 {\usepackage{color}}{}

\newcounter{bla}

\journal{Computer Physics Communications}

\makeatother

\usepackage{babel}

\begin{document}

\title{MEKS: a program for computation \\
of inclusive jet cross sections at hadron colliders}

\author[1]{Jun Gao\corref{jun}}

\author[1]{Zhihua Liang}

\author[2]{Davison E. Soper}

\author[3]{Hung-Liang Lai}

\author[1]{Pavel M. Nadolsky}

\author[4,5]{C.-P. Yuan}

\cortext[jun]{Corresponding author.\\
 \textit{E-mail address:} jung@smu.edu}

\address[1]{Department of Physics, Southern Methodist University, Dallas, TX
75275-0175, USA}

\address[2]{Institute of Theoretical Science, University of Oregon, Eugene,
OR 97403-5203, USA}

\address[3]{Taipei Municipal University of Education, Taipei, Taiwan}

\address[4]{Department of Physics and Astronomy, Michigan State University,
East Lansing, MI 48824-1116, U.S.A.}

\address[5]{Center for High Energy Physics, Peking University, Beijing 100871,
China}

\begin{abstract}
EKS is a numerical program that predicts differential cross sections for
production of single-inclusive hadronic jets and jet pairs at
next-to-leading order (NLO) accuracy in a perturbative QCD calculation.
We describe MEKS 1.0, an upgraded EKS program with increased numerical precision,
suitable for comparisons to the latest experimental data from
the Large Hadron Collider and Tevatron. The program integrates the
regularized patron-level matrix elements over the kinematical phase space
for production of two and three partons using the VEGAS algorithm.
It stores the generated weighted events in finely binned
two-dimensional histograms for fast offline analysis. A user
interface allows one to customize computation of inclusive jet
observables. Results of a benchmark comparison of the MEKS program
and the commonly used FastNLO program are also documented. \end{abstract}

\maketitle

\noindent \textbf{PROGRAM SUMMARY}

\begin{small}

\noindent {\em Manuscript Title:} MEKS 1.0: a program for computation
of inclusive jet cross sections at hadron colliders \\
 {\em Authors:} Jun Gao, Zhihua Liang, Davison E. Soper, Hung-Liang
Lai, Pavel M. Nadolsky, C.-P. Yuan\\
 {\em Program Title:} MEKS 1.0\\
 {\em Journal Reference:} \\
{\em Catalogue identifier:} \\
 {\em Preprint number:}
 SMU-HEP-12-10\\
{\em Licensing provisions:} none\\
{\em Programming language:} Fortran (main program), C (CUBA
library and analysis program).\\
 {\em Computer:} all\\
{\em Operating system:} any UNIX-like system\\
{\em RAM:} $\sim$ 300 MB\\
{\em Supplementary material:} 
{\em Classification:} 11.1\\
{\em Nature of problem:} Computation of differential cross sections
for inclusive production of single hadronic jets and jet pairs at
next-to-leading order accuracy in perturbative quantum chromodynamics.
\\
 {\em Solution method:} Upon subtraction of infrared singularities,
the hard-scattering matrix elements are integrated over available
phase space using an optimized VEGAS algorithm. Weighted events are
generated and filled into a finely binned two-dimensional histogram,
from which the final cross sections with typical experimental binning
and cuts are computed by an independent analysis program. Monte Carlo
sampling of event weights is tuned automatically to get better efficiency.
\\
 {\em Running time:} Depends on details of the calculation and
sought numerical accuracy. See benchmark performance in Section~\ref{sec:Benchmarks}.
\\
\end{small}

\section{Introduction}

Production of particle jets in high-energy collisions is a cornerstone
process of the physics program at the CERN Large Hadron Collider (LHC)
and Fermilab Tevatron $p\bar{p}$ collider. Historically, observation
of final-state jets formed by hadrons in $e^{+}e^{-}$ collisions
confirmed the asymptotic freedom of strong interactions. In modern
experiments, measurements of inclusive jet production at $pp$ and
$p\bar{p}$ colliders have reached unprecedented precision. They serve
both for exacting tests of perturbative quantum chromodynamics (PQCD)
and for searches for hypothetical new interactions at the highest
energy scales attained. Within PQCD, measurements of single-jet production
cross sections at the Tevatron Run-2 constrain the QCD coupling constant~\cite{Abazov:2009nc}
and parton distribution functions (PDF) in the proton~\cite{Lai:2010vv,Guzzi:2011sv,Martin:2009iq,Ball:2010de}.
Jet production has unique sensitivity to the momentum distribution
of gluons with large momentum fractions $x$, which is not available
in other scattering processes~\cite{JetCorrelations2012}. Invariant
mass distributions of dijets~\cite{Chatrchyan:2011ns}, dijet angular
distributions~\cite{Khachatryan:2011as,Aad:2011aj}, and other jet
observables at the LHC~\cite{:2010wv,Chatrchyan:2011qta,CMS:2011ab}
are examined to search for quark compositeness and heavy particle
resonances. All these analyses depend on reliable theoretical computations
that continue to evolve to stay on par with experimental developments.

From the experimental point of view, jet production has an advantage
of very high statistics and a drawback of sizable systematic errors
associated with the complexities of jet reconstruction. On the theory side,
predictions for jet observables remain known to next-to-leading order
(NLO) only~\cite{Ellis:1992en,Kunszt:1992tn,Nagy:2001fj,Nagy:2003tz}.
Theoretical uncertainties due to the QCD scale dependence and the fixed-order
model for the jet algorithm are comparable to the experimental errors.
Some phenomenological studies also include partial next-to-next-to-leading
order contributions to jet cross sections obtained by threshold resummation
\cite{Kidonakis:2000gi}.

This paper describes MEKS, a program for predicting probabilities of
observation of a single jet or of jet pairs that are accompanied by
arbitrary final states,
$p+p\!\!\!{}^{^{(-)}}\rightarrow\mbox{jet}+X$ and
$p+p\!\!\!{}^{^{(-)}}\rightarrow\mbox{jet}+\mbox{jet}+X$. An early
numerical code (EKS) for the NLO calculation of single inclusive jet
and dijet distributions was developed by S. D. Ellis, Z. Kunszt and
D. E. Soper in the 1990's~\cite{Ellis:1992en} based on the subtraction
method. The MEKS program is based on the original EKS calculation
that has been augmented by new elements to boost its stability,
flexibility, and efficiency. Two other widely used numerical
programs are NLOJET++~\cite{Nagy:2001fj,Nagy:2003tz} -- a complete
calculation of inclusive one and two jet cross sections at
NLO, -- and FastNLO~\cite{Kluge:2006xs,ftnlo:2010xy}, which
provides a fast interpolation of NLOJET++ cross sections in the
kinematical bins of already published experimental
measurements.\footnote{NLOJET++ can also be used to calculate three jet
inclusive cross sections, but that feature is not relevant for this
paper.}
Besides these \textit{fixed-order} calculations, POWHEG combines the
NLO jet production cross sections with \textit{leading-logarithm}
QCD showering effects~\cite{Alioli:2010xa}.

Precision calculations for these processes are challenging because
of the rapid falloff of the cross sections with the jet's $p_{T}$
and rapidity. In a typical experimental data set, jet cross sections
vary by up to 6-9 orders of magnitude. In addition, large numerical
cancelations occur between some $2\rightarrow2$ and $2\rightarrow3$
contributions due to the presence of QCD singularities. MEKS
undertakes several measures to handle these issues and to achieve
relative accuracy of order one percent in the numerical simulations.
The MEKS output is produced in the form of two-dimensional
differential cross sections ($d^{2}\sigma/(dp_{T}\, dy)$,
$d^{2}\sigma/(dm_{jj}\, dy)$, ...) after integration over the
unobserved momentum components using the VEGAS method from the
CUBA2.1 library~\cite{Hahn:2004fe}. The Monte-Carlo integration is
automatically optimized to improve the speed of the calculation. The
generated events are written into finely binned two-dimensional
histograms that can be rebinned into any set of coarse bins of a
given experiment at the stage of the user's final analysis. This
format is different from the FastNLO format, which provides the
cross sections in the coarse bins that are taken from pre-existing
experimental publications. Residual theoretical uncertainties in
such an NLO calculation are currently comparable to typical
experimental errors. Thus, together with the MEKS code, we present a
detailed benchmark comparison with the independent FastNLO code and
comment on the stability of the NLO jet calculations with respect to
the choice of QCD scales.

This document is structured as follows. Section~\ref{sec:BasicTheory}
reviews theoretical prerequisites for the calculation of single inclusive
jet and dijet cross sections at NLO. Section~\ref{sec:Description-of-MEKS}
describes the structure of the program, its inputs and outputs, installation,
and running. Section~\ref{sec:Benchmarks} considers the performance
of the program and summarizes its benchmark comparison against FastNLO.
Section~\ref{sec:Conclusion} contains the conclusion.

\section{Production of hadronic jets at NLO in perturbative QCD \label{sec:BasicTheory}}

\subsection{Factorized cross sections and measurement functions}

A jet reveals itself by tracks and calorimeter energy depositions
left by final-state hadrons in a collider detector. Numerous
particles comprise a typical jet, and their detailed distribution is
complicated. Nevertheless the probability for producing the whole
jet can be deduced with high confidence from the cross section for
production of the partons (quarks or gluons) that are the jet's
progenitors. Each contributing parton-level cross section can be
computed in PQCD. It is included into the total cross section for
producing the jet according to the jet algorithm, \emph{i.e.}, the
convention adopted to define the jet in terms of the four-momenta of
its constituent particles. The parton-level prediction must be
corrected for effects of final-state showering, hadronization, and
event pile-up in order to be compared to the observables recorded by
the detector.

PQCD provides a generic cross section for production of $N$ partons
in scattering of hadrons $H_{1}$ and $H_{2}$ with center-of-mass
energy $\sqrt{s}$, \begin{align}
\frac{d\sigma(s)}{d\Phi_{N}}= & \sum_{a_{1},a_{2}}\int_{0}^{1}d\xi_{1}\int_{0}^{1}d\xi_{2}f_{a_{1}/H_{1}}(\xi_{1},\alpha_{s},\mu_{F})\, f_{a_{2}/H_{2}}(\xi_{2},\alpha_{s},\mu_{F})\nonumber \\
 & \times\frac{d\widehat{\sigma}_{a_{1}a_{2}}(\xi_{1}\xi_{2}s;\,\alpha_{s},\,\mu_{R},\,\mu_{F})}{d\Phi_{N}}.\end{align}
Here
$f_{a_{1}/H_{1}}(\xi_{1},\alpha_{s},\mu_{F})$ and
$f_{a_{2}/H_{2}}(\xi_{2},\alpha_{s},\mu_{F})$ are the parton
distribution functions (PDFs) of intermediate partons $a_{1}$ and
$a_{2}$ in the parent nucleons $H_{1}$ and $H_{2}$. The factor
$\widehat{\sigma}_{a_{1}a_{2}}$ is the cross section for hard
scattering of $a_{1}$ and $a_{2}$, which is computable as a series
in the running coupling strength $\alpha_{s}(\mu_R)$. The parents'
momentum fractions carried by $a_{1}$ and $a_{2}$ are $\xi_{1}$ and
$\xi_{2}$, while $\mu_{R}$ and $\mu_{F}$ are the renormalization
scale and factorization scale that arise in $\alpha_{s}(\mu_{R})$
and the PDFs, respectively.

$\Phi_{N}(p_{1},p_{2},...,p_{N})$ is the phase space for production
of $N$ (massless) partons with four-momenta $p_{i}$. To relate this
parton-level cross section to a jet observable $I$ that can be
measured, such as the cross section in some experimental bins, it
must be folded into an integral \begin{equation}
I=\sum_{N=2}^{\infty}\:\sum_{f.s.c.}\int
d\Phi_{N}(p_{1},..,p_{N})\,\frac{d\sigma}{d\Phi_{N}}\,
S_{N}(p_{1},...,p_{N}).\label{I}\end{equation} The function
$S_{N}(p_{1},...,p_{N})$ represents constraints on the partons'
momenta imposed by various steps of the measurement, including the
jet algorithm. The right-hand side of Eq.~(\ref{I}) is summed over
all final-state configurations (f.s.c.) and parton types that may
contribute. The number $N$ of the final states is summed from $2$ to
infinity, but large-$N$ configurations are suppressed by a high
power $\alpha_{s}^{N}$ of the small parameter $\alpha_{s}.$

A next-to-leading-order (NLO) calculation of the single-inclusive
jet or dijet cross section includes only $2\rightarrow2$ and
$2\rightarrow3$ parton scattering contributions. At NLO, the jet observable
is represented in the calculation by two functions,
$S_{2}(p_{1},p_{2})$ and $S_{3}(p_{1},p_{2},p_{3})$. The expectation
of the observable is
\begin{equation}
\begin{split}{\cal I}= & \int\! dy_{1}\, dp_{2}\, dy_{2}\, d\phi_{2}\ \frac{d\sigma}{dy_{1}\, dp_{2}\, dy_{2}\, d\phi_{2}}\ S_{2}(p_{1},p_{2})\\
 & +\int\! dy_{1}\, dp_{2}\, dy_{2}\, d\phi_{2}\, dp_{3}\, dy_{3}\, d\phi_{3}\ \frac{d\sigma}{dy_{1}\, dp_{2}\, dy_{2}\, d\phi_{2}\, dp_{3}\, dy_{3}\, d\phi_{3}}\ S_{3}(p_{1},p_{2},p_{3})\;\;.\end{split}
\label{eq:NLOstart}\end{equation}
 The cross sections $\sigma$ depend on $p_{1},p_{2}$ for the case
of two final state partons and on $p_{1},p_{2},p_{3}$ for the case
of three final state partons. The parton momenta are determined by
the absolute value of the transverse momentum, $p_{i}$, the rapidity,
$y_{i}$, and the azimuthal angle, $\phi_{i}$. Momentum conservation
has been applied to reduce the number of independent integration variables.
The functions $S_{3}$ and $S_{2}$ are symmetric under interchange
of the parton labels 1,2,3 and must be related by infrared safety
conditions: $S_{3}(p_{1},p_{2},p_{3})$ must approach $S_{2}(p_{1},p_{2}+p_{3})$
when the parton momenta $p_{2}$ and $p_{3}$ become collinear with
each other, or when $p_{3}$ tends to zero. In Eq.~(\ref{eq:NLOstart}),
both terms are infrared divergent and need to be regulated by working
in $4-2\epsilon$ space-time dimensions (with $\epsilon\rightarrow0$)
instead of 4 dimensions.%
\footnote{The needed modifications to the $(4-2\epsilon)$-dimensional integration
measure are suppressed in Eq.~(\ref{eq:NLOstart}).%
} For this reason, some manipulation is needed to bring ${\cal I}$
into a form in which the integrals can be calculated by Monte Carlo
numerical integration.

In this paper, we are concerned with double-differential jet cross
sections. That is, there are two variables $A$ and $B$ that are
functions of the parton momenta: $A_{3}(p_{1},p_{2},p_{3})$ and $B_{3}(p_{1},p_{2},p_{3})$,
or $A_{2}(p_{1},p_{2})$ and $B_{2}(p_{1},p_{2})$. These functions
are infrared-safe in the sense defined above. For instance, $A$ might
be the invariant mass of the leading two jets in an event, and $B$
might be the difference in the rapidities of the two jets, where the
jets are defined, for instance, by the $k_{T}$ jet algorithm. Then
the functions $S$ in Eq.~(\ref{eq:NLOstart}) specify whether the
event contributes to a certain bin in a histogram of the cross section:
\begin{equation}
\begin{split}S_{2}(p_{1},p_{2})={} & \theta(|A_{2}(p_{1},p_{2})-a|<\Delta_{A})\,\theta(|B_{2}(p_{1},p_{2})-b|<\Delta_{B})\;\;,\\
S_{3}(p_{1},p_{2},p_{3})={} & \theta(|A_{3}(p_{1},p_{2},p_{3})-a|<\Delta_{A})\,\theta(|B_{3}(p_{1},p_{2},p_{3})-b|<\Delta_{B})\;\;.\end{split}
\end{equation}

The subtraction method for calculating ${\cal I}$ is explained in
some detail in Ref.~\cite{Kunszt:1992tn}. There is no point in
repeating this explanation here. It may be helpful, however, to illustrate
the final form of the Monte Carlo integration in a simplified model
calculation that is independent of rapidities.
In the simplified model, we integrate over a single one-dimensional
variable $p_{2}$ in the first term and over two one-dimensional variables
$p_{2}$ and $p_{3}$ in the second term.
After applying the
subtraction method to tame divergences, the model integral has the
form \begin{equation}
\begin{split}{\cal I}={} & \int_{0}^{\infty}\! dp_{2}\ f_{2}(p_{2})\,\theta(|A_{2}(p_{2})-a|<\Delta_{A})\\
 & +\int_{0}^{\infty}\! dp_{2}\int_{0}^{\infty}\! dp_{3}\ \bigg[\frac{f_{3}(p_{2},p_{3})}{p_{3}}\,\theta(|A_{3}(p_{2},p_{3})-a|<\Delta_{A})\\
 & \quad-\frac{f_{3}(p_{2},0)\theta(p_{3}<p_{2})}{p_{3}}\,\theta(|A_{2}(p_{2})-a|<\Delta_{A})\bigg]\;\;.\end{split}
\label{eq:NLO1}\end{equation}
 There is a divergence from the first term of the second integral arising
from the region $p_{3}\to0$. However, in this model, the infrared
safety property of the jet definition is $A_{3}(p_{2},0)=A_{2}(p_{2})$.
Then the divergence is canceled because of the subtraction term. The
integral as written is suitable for calculation by Monte Carlo numerical
integration.

The program described here is modified from the original EKS program,
which used an elaborate method to calculate a double differential
jet cross section, because the computer power available when the program
was written was quite limited. The method used here is to calculate
the jet observables $A$ and $B$ for each point in the Monte Carlo
integration and put them into very small bins. Then a separate analysis
routine can calculate the desired cross section in any larger bin
desired. In our model calculation, this means that we calculate a
collection of integrals \begin{equation}
\begin{split}{\cal I}_{i}={} & \int_{0}^{\infty}\! dp_{2}\ f_{2}(p_{2})\,\theta(|A_{2}(p_{2})-a_{i}|<\Delta_{A})\\
 & +\int_{0}^{\infty}\! dp_{2}\int_{0}^{\infty}\! dp_{3}\ \bigg[\frac{f_{3}(p_{2},p_{3})}{p_{3}}\,\theta(|A_{3}(p_{2},p_{3})-a_{i}|<\delta_{A})\\
 & \quad-\frac{f_{3}(p_{2},0)\theta(p_{3}<p_{2})}{p_{3}}\,\theta(|A_{2}(p_{2})-a_{i}|<\delta_{A})\bigg]\;\;.\end{split}
\label{eq:NLO2}\end{equation}
 Here the bin widths $\delta_{A}$ are very small. From this information,
the integral over a larger bin $|A-a|<\Delta_{A}$ can be calculated
later for any $a$ and $\Delta_{A}$ desired. In the program described
here, the Monte Carlo points $p_{2}$ and $p_{3}$ are chosen using
the VEGAS algorithm \cite{Lepage:1977sw} with some modifications.

\subsection{Main theoretical inputs}

Theoretical inputs must be selected carefully in the comparisons of
NLO cross sections, since they may cause non-negligible differences
in predictions.
\begin{itemize}
\item \textbf{Jet algorithm.} When calculating the distribution of jet observables,
we need to use the same jet algorithms as the ones in the
experimental measurements. In a PQCD calculation, the experimental
jet algorithm is approximated by a measurement function
$S_N(p_1,...,p_N)$ that acts on a small number of partons. (See the
previous section.) The cone-based Midpoint
algorithm~\cite{Blazey:2000qt} is most frequently used at the
Tevatron, while the cluster-based anti-$k_{T}$
algorithm~\cite{Cacciari:2008gp} is in standard use at the LHC. The
only difference between the Midpoint algorithm and modified Snowmass
algorithm~\cite{Blazey:2000qt} used in the original EKS program is
that the Midpoint algorithm always uses the midpoint of the two partons'
directions as one of the possible seeds for a new protojet.
In the NLO theoretical
calculations for single-jet or dijet production that include at most
three final-state partons, the cluster-based
$k_{T}$~\cite{Catani:1993hr,Ellis:1993tq}, anti-$k_{T}$, and Cambridge-Aachen
(CA)~\cite{Dokshitzer:1997in} algorithms are equivalent. The
Midpoint algorithm
is generally different from these algorithms.
\item \textbf{The recombination scheme} is a procedure for merging two nearby
partons into one jet. For example, the 4D scheme computes the jet's
4-momentum by adding the 4-momenta of the jet's constituents. The
$E_{T}$ scheme finds the momentum of the merged jet by adding the
scalar $E_{T}$ values, then averaging over the partons' directions
with the statistical weights given by the individual $E_{T}$ values
\cite{Salam:2009jx}. The 4D scheme is often adopted by the recent
experiments at both the Tevatron and LHC. Different choices of the
recombination scheme can cause differences in the NLO predictions,
as will be shown later in the benchmark comparison section. Note that,
with the 4D scheme, the jet could be massive, which means that the
jet's pseudorapidity will not be equal to its rapidity.
\item \textbf{The jet acceptance} conditions specify if the jet will be
considered in the final observables. For example, to be taken into
account, the total transverse momentum $p_{T}$ of the jet in the
acceptance region must be larger than a threshold value
$p_{T}^{min}$ of a few tens GeV/c. In NLO calculations of
single-inclusive jet distributions, the jet acceptance conditions
practically have no effects. In dijet production, they may change
the cross sections by small amounts by affecting the selection of
two leading jets.
\item \textbf{Renormalization and factorization scales.} The scale choice
is only related to theory and has no correspondence in experiment.
It is conventional to choose the renormalization and factorization
scales to be of the order of the typical transverse momentum $p_{T}$
of the jet(s): $\mu_{R}\sim\mu_{F}\sim p_{T}$. In contributions with
two resolved jets, $p_{T}$ naturally corresponds to the transverse
momentum of either of the final-state jets (which are about equal
by momentum conservation). More ambiguity is present in contributions
with three resolved jets, when $p_{T}$ can correspond to the transverse
momentum of either of the jets in each event or to a combination of
three transverse momenta. The conventional choices made by experimentalists
are to set both of the scales to individual jet $p_{T}$ for single
inclusive jet production and to the average $p_{T}$ of the two leading
jets for dijet production.
\end{itemize}

\subsection{Typical kinematical variables \label{sub:Typical-kinematical-distributions}}

For completeness, we provide definitions of typical kinematical variables
arising in recent measurements. The \emph{leading jet} has the largest
transverse momentum $p_{Tj}$ among all jets in each scattering event.
\textbf{Single-inclusive jet} cross sections may refer either to distributions
of the leading jets or individual jets (so that all jets hitting a
particular kinematic bin in every event are counted). Currently, MEKS
has a built-in mode for computing single-inclusive distributions of individual jets,
although distributions for the leading jets can be easily implemented. The
transverse momentum and rapidity of the individual jet are denoted
by $p_{Tj}$ and $y_{j}$, respectively.

In\textbf{ dijet production, }kinematical variables are formed from
the four-momenta of the first and second leading jets,
$p_{1j}^{\mu}$ and $p_{2j}^{\mu}$. Common distributions are given in
terms of the dijet's invariant mass
$m_{jj}=\sqrt{(p_{1j}+p_{2j})^{2}}$, the maximal absolute rapidity
value
$y_{max}=\max\left(\left|y_{1j}\right|,\left|y_{2j}\right|\right)$
in the jet pair, and the angular variable
$\chi=\exp(|y_{1j}-y_{2j}|)$.

\section{Description of the program \label{sec:Description-of-MEKS}}

\subsection{Algorithm}

\begin{figure}[h!]
\centering \includegraphics[width=1\textwidth]{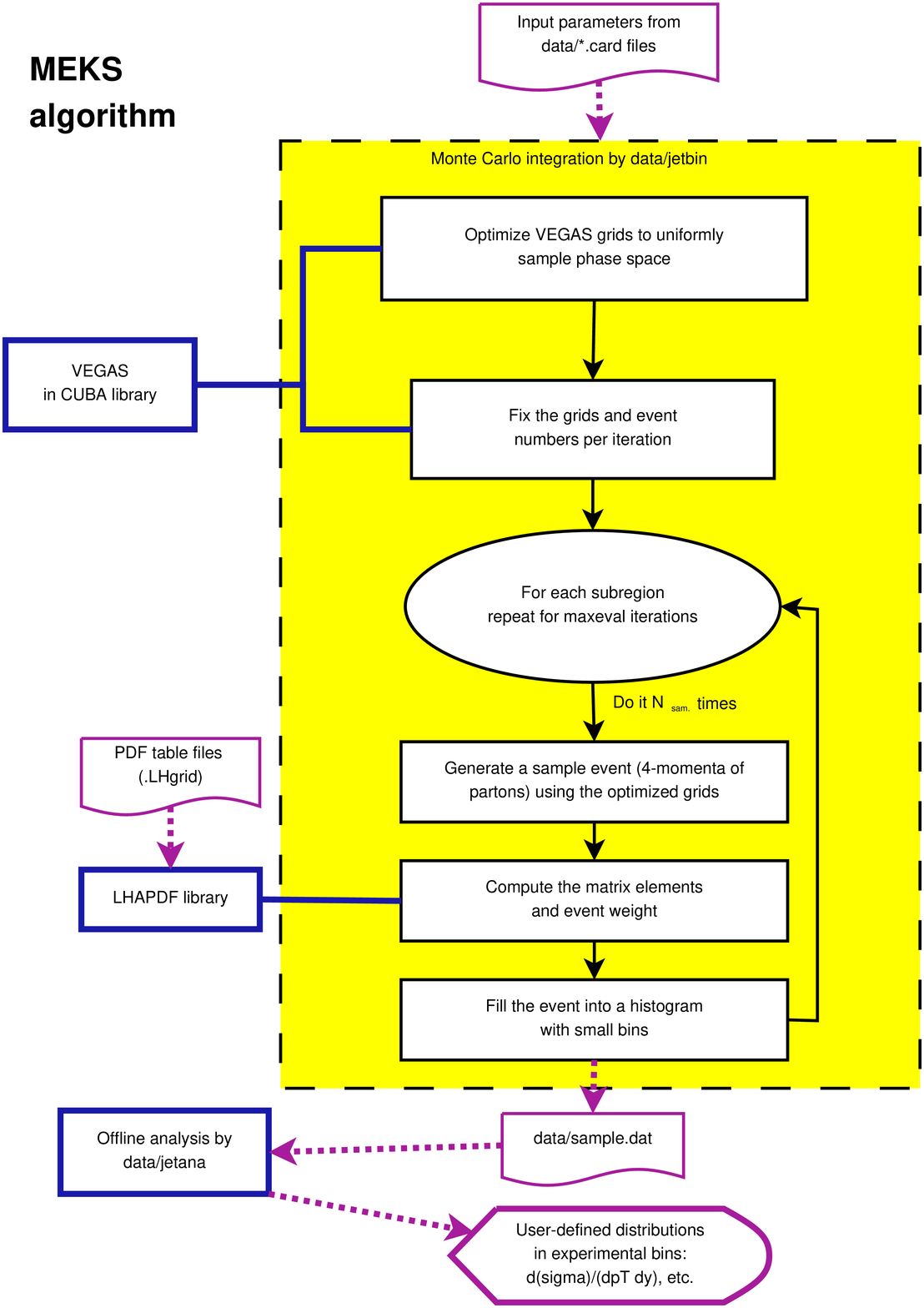}
\caption{The algorithm of the computation.}

\label{flowchart}
\end{figure}

The goal of the MEKS program is to calculate double-differential
cross sections for single-inclusive jet or dijet production at
hadron colliders up to NLO in QCD. Its basic algorithm is shown in
Fig.~\ref{flowchart}. All executables, input files, and output files
are stored in the subdirectory \texttt{data/}. The main computation
is carried out in an executable \texttt{jetbin}, which performs
Monte-Carlo integration of fully differential NLO cross sections
using the VEGAS algorithm provided by the CUBA
library~\cite{Hahn:2004fe}. The input parameters are read from
several \texttt{.card} files. In view of the rapid falloff of the
cross section across the available phase space region, the integration
volume is divided into subregions which are handled separately, and
additional optimization is performed to achieve percent-level
accuracy in each subregion. The integration runs in a sequence of
three steps. First, it scans over entire kinematic regions of the
jet observables and calculates Monte Carlo sampling weights for
different regions in order to improve the efficiency of the
sampling. In step 2, the program generates and optimizes the VEGAS
grids for MC samplings using the above weights. In the final step,
it performs Monte Carlo sampling based on the corresponding VEGAS
grids and generates the weighted events. For each sampling event,
the program generates independent components of partonic momenta
(4
components in $2\rightarrow2$ subprocesses and 7 components in
$2\rightarrow3$ subprocesses). It then computes the cross section
weight for the event as described in Sec.~\ref{sec:BasicTheory},
using the parton-level cross sections, PDFs, and measurement
functions. Parametrizations of the PDFs are provided by the LHAPDF
library \cite{LHAPDF}, which requires PDF table files (\texttt{.LHgrid})
as an input for interpolation. The momentum components and
final weight of each event are written into auxiliary
two-dimensional histograms in an ASCII file. The bin sizes are
chosen to be substantially smaller than in a typical experiment, of
order 1 GeV for momentum variables, and 0.1 for rapidity variables.
Finally, when the computation is finished, the auxiliary histograms
can be rebinned into the bins of a selected experiment using the
\texttt{jetana} program. %

\subsection{Installation \label{sub:Installation}}

The MEKS code is installed by unpacking the archive (.zip) file and
running \texttt{make} in the main directory. An external LHAPDF library
is required and the path of the LHAPDF library should be set in
the \texttt{makefile} before running \texttt{make}. The compilation
requires \texttt{gcc}, \texttt{g++}, and \texttt{gfortran}
compilers. The compiler will generate two executables in the
subdirectory \texttt{data}: \texttt{jetbin} for generating
the intermediate histogram files, and \texttt{jetana} for
calculating the final differential cross sections. Running
\texttt{make clean} will delete all the compiled libraries and
executables.

\subsection{The source \label{sub:Source}}

The main directory contains 8 source files together with 3
subdirectories. \texttt{jetbin.f} contains the main program for the
Monte Carlo sampling. The EKS function \texttt{RENO()} that returns
the integral for the jet cross sections is stored in
\texttt{ekscode.f}. The module \texttt{modsub.f} contains
subroutines for reading the inputs, clustering and selection of
jets, recording of the events, generation of the histograms, and
interface to the CUBA library. \texttt{lha\_interface.f} is the
interface to the LHAPDF library. \texttt{user.inc} and
\texttt{var.inc} are header files with definitions of variables and
common blocks. \texttt{jetana.cxx} is a C++ source file for the
offline rebinning of the histogram files into the final
double-differential cross sections. These files are not supposed to
be changed by the user. Customization of inputs and outputs can be
done by editing the module \texttt{userfunction.f}, as described in
Sec.~\ref{sub:Customization}.

Besides the subdirectory \texttt{data/} described above, the
subdirectory \texttt{lib/} contains a simplified CUBA library. It is
written in C and includes only VEGAS algorithm. Examples in the
subdirectory \texttt{example/} show how to set the input files for
representative Tevatron and LHC measurements.

\subsection{Input parameters \label{sub:Input-parameters}}

\begin{table}
\begin{centering}
\begin{tabular}{|c|c|c|}
\hline \texttt{promode} & Observable & Distribution\tabularnewline
\hline \hline 1 & Single-inclusive &
$d^{2}\sigma/(d\left|y_{j}\right|dp_{Tj})$\tabularnewline \hline 2 &
Dijet & $d^{2}\sigma/(dy_{max}dm_{jj})$\tabularnewline \hline 3 &
Dijet & $d^{2}\sigma/(d\chi dm_{jj})$\tabularnewline \hline 4 &
User-specified & $d^{2}\sigma/(dv_1 dv_2)$\tabularnewline \hline
\end{tabular}
\par\end{centering}

\caption{Implemented computational modes.}

\end{table}

The input parameters are specified in the files with the extension
\texttt{.card}. Each line contains a record for one input variable:
a character tag with the name of the variable, followed by the
variable's value.

\texttt{proinput.card} specifies overall controls for the
computation
\begin{itemize}
\item \texttt{pdfname} is the name of the PDF file to be used in the calculation, \emph{e.g.},
\texttt{cteq66.LHgrid} if CTEQ6.6 central PDFs are used.
\item \texttt{pdfmem} specifies which member of the PDF set from the LHAPDF
library is to be invoked. For example, set \texttt{pdfmem=0} to
access the central PDF set or another integer to access an error PDF
according to the definition provided by LHAPDF.
\item \texttt{promode} specifies the double-differential distribution. Set
\texttt{promode=1,2, or 3} to produce common distributions listed in
Table 1, where the kinematical variables are defined in
Sec.~\ref{sub:Typical-kinematical-distributions}. Alternatively,
choose \texttt{promode=4} to compute a user-defined distribution
that is specified in \texttt{userfunction.f.}
\item \texttt{smode} specifies the choice of the hard momentum scale $\mu_{0}$
that defines the central value of the factorization and
renormalization scales: $\mu_{F}=\kappa_{F}\mu_{0}$ and
$\mu_{R}=\kappa_{R}\mu_{0}$, where $\kappa_{F}$ and $\kappa_{R}$ are
the prefactors of order unity that are input as \texttt{fscale} and
\texttt{rscale} later in the card. Currently \texttt{smode = 1} sets
the central scale to the individual jet $p_{T}$ for single-inclusive
jet production and to the average $p_{T}$ of two leading jets for
dijet production. \texttt{smode = 2} sets the central scales equal
to the leading jet's $p_{T}$ for both single-inclusive jet and dijet
production. For the user-specified calculation mode
(\texttt{promode=4}), the central scale should be set directly in
\texttt{userfunction.f} by the user.
\item \texttt{loop} specifies the order of the QCD coupling: \texttt{0}
for a LO calculation and \texttt{1} for a full NLO calculation (LO
+ NLO corrections).
\item \texttt{ppcollider} represents the type of the collider: \texttt{0} for
a $p\bar{p}$ machine and \texttt{1} for a $pp$ machine.
\item \texttt{sqrtSS} gives the center-of-mass energy $\sqrt{s}$ of the collider in
GeV.
\item \texttt{fscale} and \texttt{rscale} are the prefactors for $\mu_{F}$
and $\mu_{R}$, cf. the discussion above.
\item \texttt{iseed} specifies the random-number seed used by the VEGAS
subroutine. If \texttt{iseed = 0} (recommended), the seeds are
generated randomly from the internal clock readings. The results
from different runs will be statistically independent and can be
combined to improve the numerical accuracy. Or, the user can specify
\texttt{iseed $\neq$ 0} to reproduce the same pseudorandom numbers
in subsequent runs.
\item \texttt{maxeval} specifies the total number of iterations carried out
in the calculation as shown in Fig.~\ref{flowchart}.
\texttt{maxeval} controls both the CPU time cost and numerical
accuracy of each job. Note that for all the calculation modes except
the user-specified one, the numbers ($N_{sam}$) of Monte Carlo
sampling points for each iteration and subregion are determined
automatically, and are not supposed to be specified by the user.
They may differ for different subregions. For example, more sampling
points can be placed into a subregion with a large volume or close
to the edges of the phase space. Since the offline analysis code can
combine results from several independent jobs to improve the
numerical accuracy, we suggest that the user runs multiple jobs on a
cluster or multi-core machine with \texttt{maxeval<5,} rather than a
single job with a large \texttt{maxeval} value.
\end{itemize}
\texttt{kininput.card} contains parameters for clustering and selection
of jets, in the same format as in \texttt{proinput.card}.
\begin{itemize}
\item \texttt{jetalgo} specifies the jet algorithm used in the calculation:
\texttt{1} for the anti-$k_{T}$ jet algorithm, \texttt{2} for the
modified Snowmass algorithm, in which the Midpoint algorithm is a
special case of \texttt{Rsep = 2} at NLO.
\item \texttt{recscheme} sets the recombination scheme used in jet clustering:
\texttt{1} for the 4D scheme and \texttt{2} for the $E_{T}$ scheme.
\item \texttt{Rcone} sets the cone size or distance parameter for the jet
algorithm used.
\item \texttt{Rsep} sets the separation parameter used in the modified Snowmass
algorithm. Note that in the theoretical calculations here, the modified
Snowmass algorithm with \texttt{Rsep = 2} is equivalent to the Midpoint
algorithm.
\item \texttt{ptcut} and \texttt{ycut} specify conditions for jet acceptance,
\emph{i.e.}, the lowest value of $p_{Tj}$ and largest absolute
value $\left|y_{j}\right|$ of the rapidity that a jet can have in
order to be taken into account in jet observables.
\item \texttt{yboost} is only active for \texttt{promode = 3}. It specifies
the upper limit on the rapidity of the dijet system $y_{b}=|y_{1}+y_{2}|/2$.
\end{itemize}
\texttt{user.card} contains additional parameters that are needed
by the user-specified mode, which means it is only necessary for \texttt{promode
= 4}. See Sec.~\ref{sub:Customization} for additional details.

In order to balance the numerical accuracy in different kinematic
regions, thus improve the efficiency, the program can divide the
entire integration volume into several (at most 10) subvolumes,
which will be calculated separately. Boundaries of each volume are
set in \texttt{liminput.card}, which starts with a comment
\texttt{region} in the first line, followed by the lower and upper
limits for the first (dimensionless) jet observable (rapidity or
$\chi$) in the next line, and then by the limits for the second jet
observable ($p_{Tj}$ or $m_{jj}$ in GeV) in the third line.

The last input file \texttt{bins.in} specifies the settings for the
offline rebinning of the final two-dimensional differential cross
sections in \texttt{jetana}. It is different from
\texttt{liminput.card,} which only specifies the optimal subvolumes
at the integration stage. Consider the following example, in which
the single-inclusive cross sections are redistributed into 2 bins of
rapidity, $0\leq y_{j}\leq1$ and \texttt{$1\leq y_{j}\leq2$}, with
the \texttt{$0\leq y_{j}\leq1$} bin containing 3 $p_{Tj}$ bins
(\texttt{30-50}, \texttt{50-80}, \texttt{80-100} GeV), and the
\texttt{$1\leq y_{j}\leq2$} bin containing 2 $p_{T}$ bins
(\texttt{40-80} and \texttt{80-100} GeV). This is achieved with the
following \texttt{bins.in:}

\renewcommand{\baselinestretch}{0.78}
\begin{verbatim}
 1     ## User's comments
 2     ## User's comments
 3     ## User's comments
 4     1                  #promode
 5     2                  #number of bins of variable 1
 6     0.0  1.0  2.0      #boundaries of bins of variable 1
 7             #Specify bins of variable 2, in bin 1 of variable 1
 8     3                  #number of bins
 9     30.0               #boundaries of bins
10     50.0
11     80.0
12     100.0
13             #Specify bins of variable 2, in bin 2 of variable 1
14     2                  #number of bins
15     40.0               #boundaries of bins
16     80.0
17     100.0
\end{verbatim}Line 1-3 and 4 contain the user's comments and value of \texttt{promode}.
The rest of the file contains the number and boundaries of the
$y_{j}$ bins (lines 5-6), then of the $p_{Tj}$ bins in the first
rapidity bin (lines 8-12), and finally of the $p_{Tj}$ bins in the
second rapidity bin (lines 14-17).

To match the format of the experimental data, for \texttt{promode =
3,} the order of the two jet variable in \texttt{bins.in} is
opposite to the one in the histogram generation, i.e., $m_{jj}$ is
in the first variable, and $\chi$ is in the second one. The
subdirectory \texttt{example} contains sample files of
\texttt{liminput.card} and \texttt{bins.in} for various
distributions at the LHC and Tevatron.

\subsection{Execution \label{sub:Execution}}

To run the program, first enter the subdirectory \texttt{data}, then
run \texttt{./jetbin \$(name)}. \texttt{\$name} specifies the name
of the running job that distinguishes the outputs of different jobs.
The program will run the 3-steps sequence for each kinematic
subregion mentioned before. The output file
\texttt{sample\$(name).dat} stores a histogram with the cross
sections and integration errors, which is updated after each
iteration. Below are the first few lines of the output file from a
test run.

\renewcommand{\baselinestretch}{0.78}
\begin{verbatim}

 1  LHC     7000  user specified   anti-kt     rec(4D)     LO(pb)
 2    PDF used:  CT10.LHgrid               member:  0
 3  muf(mur): 0.1000E+01 (0.1000E+01) *user_defined dR: 0.6000E+00  Rsep:  0.2000E+01
 4  jet acceptance:  |eta|<4.40 pt> 15.00 GeV dijet boost (only for chi): |yb|<1.00
 5  2Dbins: rap/chi/v1 vs. pt/mass/v2
 6  0.10  1.00
 7      2      0    3.4
 8           6
 9  0.0000E+00  0.1000E+01  0.7000E+02  0.3310E+04
10  0.1000E+01  0.2000E+01  0.1600E+03  0.3930E+04
11  0.2000E+01  0.3000E+01  0.3700E+03  0.4640E+04
12  0.3000E+01  0.3500E+01  0.1180E+04  0.5040E+04
13  0.3500E+01  0.4000E+01  0.1760E+04  0.5470E+04
14  0.4000E+01  0.4400E+01  0.2550E+04  0.4270E+04
15    101036.549705923        17786.8878510429
16    76281.7571472343        14782.5878433955
17    90096.2647871803        17148.2526023941
18    64693.6882833766        12677.7540684543
19    70691.2937015596        13201.9918739291
20    87217.4486225131        15296.3531357402
21    58842.6156645410        10017.7794593842
22    58440.4567807920        9565.31015305604
23    43218.4690333846        7188.36482488373
24    36809.3489243188        6839.88584958308
25    53239.8950061995        8663.33778921252
26    42076.4229811524        6838.48396473720
27    53591.3648097391        7698.18602397479
28    45424.7820507153        6697.38457240572
29    33785.8434883951        5971.33017051214
30    33129.2531779327        6132.28492415771
      ...
\end{verbatim}

Lines 1-6 is a header describing the calculation. As we can see, it
is for a user-specified double-differential cross section
(\texttt{promode=4}) at the LHC with $\sqrt{s}\ =7$ TeV, at the LO
using the anti-$k_{T}$ jet algorithm and 4D recombination scheme.
The widths of the two jet observables ($|y_{j1}-y_{j2}|/2$ and
$m_{jj}$) in the fine histograms are 0.1 and 1 GeV, respectively. Line 7
contains the number of iterations that are finished (\texttt{2}),
the \texttt{iseed} value (\texttt{0}), and the elapsed CPU time in
minutes (\texttt{3.4}). Lines 8-14 shows the number of subregions
(\texttt{6}) and the boundaries of the first and second jet
observables in each subregion. The remaining lines contain the
finely-binned histogram, which lists the cross sections (left
column) and their integration errors (right column) in picobarns.
They are sorted
in the ascending order by the subregions' ID, then by the first jet
observable, and then by the second jet observable. Here the
integration errors are estimated by the standard procedure of Monte
Carlo samplings as
\begin{equation}
\delta=(\sum{\omega_i^2}-[\sum{\omega_i}]^2/N_{sam})^{1/2},
\end{equation}
where the sum is taken over all the sample points $i$ with
contribution $\omega_i$ to the cross sections in the corresponding
histogram, and $N_{sam}$ is the total number of sample points.

From this output histogram file, the final differential cross
sections are computed simply by running \texttt{./jetana
sample\$(name).dat} and using \texttt{bins.in} as the input. The
\texttt{jetana} code prints the final cross sections to both the
screen and output file \texttt{xsec.out}. Remember that the
generation of the histograms and their final analysis are
done in two separate steps: one can obtain the results for
a different binning simply by changing \texttt{bins.in,} without regenerating the
histograms. Below, we show the first few lines of \texttt{xsec.out}
from the same test run as above.

\renewcommand{\baselinestretch}{0.78}
\begin{verbatim}

LHC     7000  user specified   anti-kt     rec(4D)     LO(pb)
   PDF used:  CT10.LHgrid               member:  0
muf(mur):  0.1000E+01  (0.1000E+01)  *user_defined dR: 0.6000E+00 Rsep:  0.2000E+01
jet acceptance:  |eta|<4.40 pt> 15.00 GeV dijet boost(only for chi): |yb|<1.00

all energy in GeV and Xsecs in pb

v2 min      v2 max      Xsec/dv1/dv2        Error
--------------------------------------------
0 < v1 < 0.5
--------------------------------------------
70.00       110.00      3.858290e+05        7.372963e+03
110.00      160.00      5.187682e+04        7.097831e+02
160.00      210.00      9.767103e+03        1.363626e+02
210.00      260.00      2.716344e+03        3.836675e+01
260.00      310.00      9.441967e+02        1.350540e+01
310.00      370.00      3.744642e+02        4.957253e+00
370.00      440.00      1.378121e+02        1.750501e+00
440.00      510.00      5.512034e+01        7.158074e-01
510.00      590.00      2.281043e+01        2.850772e-01
590.00      670.00      9.896396e+00        1.283644e-01
670.00      760.00      4.507472e+00        5.775927e-02
760.00      850.00      2.141786e+00        2.761155e-02
850.00      950.00      1.043917e+00        1.314810e-02
950.00      1060.00     5.053051e-01        6.317605e-03
1060.00     1180.00     2.409892e-01        3.000414e-03
1180.00     1310.00     1.155737e-01        1.420040e-03
1310.00     1450.00     5.407524e-02        6.718418e-04
1450.00     1600.00     2.512264e-02        3.128897e-04
1600.00     1940.00     8.453138e-03        7.948315e-05
1940.00     2780.00     9.715742e-04        7.791561e-06
...
\end{verbatim}

As we mentioned earlier, the user can run multiple parallel jobs and
combine the results using the analysis code (remember to set
\texttt{iseed = 0}). The combination can be carried out in two ways.
The user can just run \texttt{./jetana sample\$(name1).dat ...
sample\$(nameN).dat} to get the combined results, or, alternatively,
run \texttt{./jetana combine sample\$(name1).dat ...
sample\$(nameN).dat} to merge the histograms
\texttt{sample\$(name1).dat ... sample\$(nameN).dat} into a new
histogram file \texttt{combine.dat}. In the latter case, run
\texttt{./jetana combine.dat} afterwards to get the combined results
just as for a single job run. Note that when combining results from
different jobs, the program uses the number of iterations as the
default weight for each job. It is equivalent to the standard
optimized procedure of using $1/\delta^2$ as weights for large
number of sampling points, since then the numerical error $\delta$
of each job is proportional to the inverse of the square root of the
iteration numbers. We choose this simple combination procedure
because it works best in the presence of statistical fluctuations
that may occur in small-sized bins.

\subsection{Customization\label{sub:Customization}}

The module \texttt{userfunction.f} is called with the
\texttt{promode=4} option. It contains two short subroutines that
control the input of the integration parameters (\texttt{readuser})
and generation of the output histograms (\texttt{userselect}). By
modifying these subroutines, the format of the input and output from
the \texttt{jetbin} code can be customized. In particular,
\texttt{readuser} reads several input parameters from
\texttt{data/user.card} that are passed into the main code through a
double-precision array \texttt{duser(10)}. The first 5 lines of
\texttt{user.card} explain what the inputs do, and they are followed
by at most ten values, with a single entry per line. These values
are kept throughout the calculation in the array \texttt{duser(10)}.
The first four values are mandatory. They represent the widths of
the histogram bins for the first and second jet observable, a
typical $p_{T}$ scale $\mu_{jet}$ (in GeV), and the number of
sampling points in each iteration (user can adjust it according to
the required numerical accuracy). $\mu_{jet}$ must be set to be of
the order of average jet $p_T$ to optimize Monte Carlo sampling. Other
parameters can be optionally defined to pass additional elements for
the array \texttt{duser(5:10)}. Instructions for the customization
are included as comments in \texttt{userfunction.f}.

\section{Performance and benchmark comparison \label{sec:Benchmarks}}

\begin{figure}[h!]
\begin{centering}
\includegraphics[width=1\textwidth]{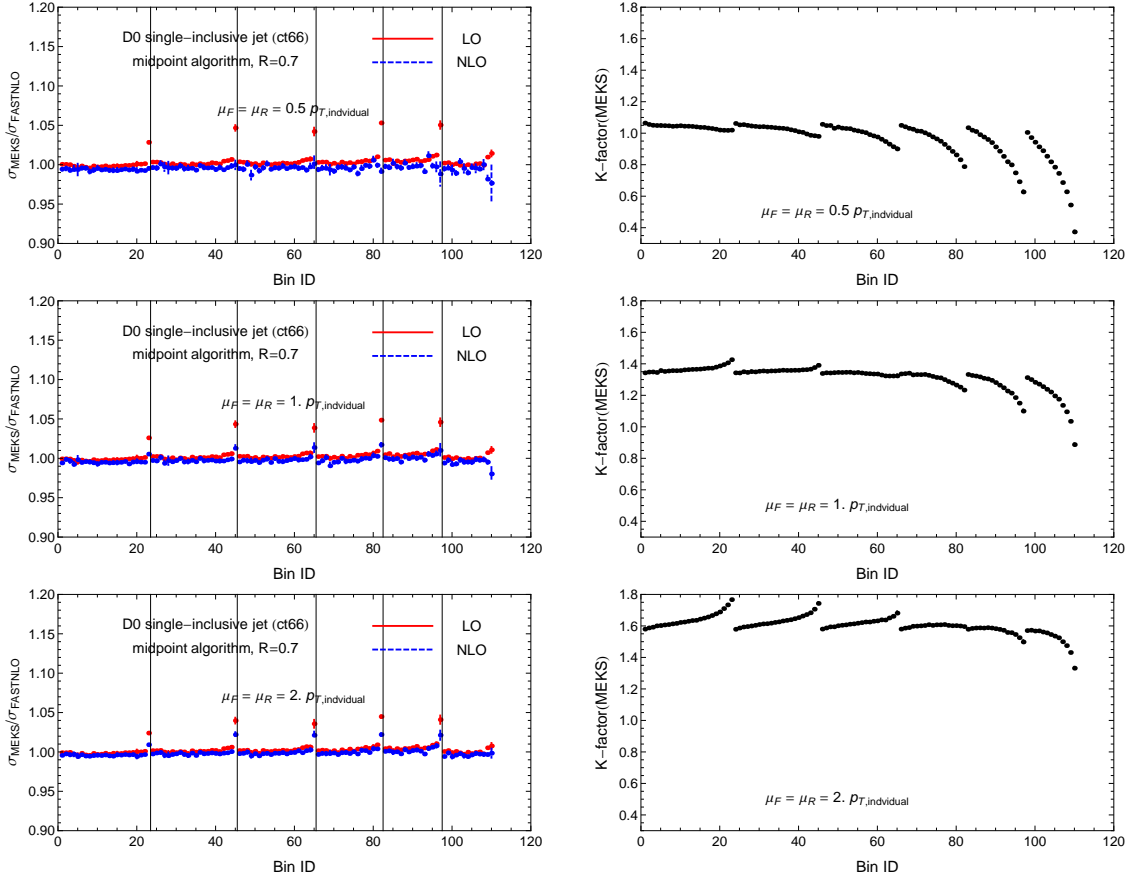}
\par\end{centering}

\caption{\label{jetcom_d0inc} Comparison of $p_{T}$ distributions
for single-inclusive jet production from MEKS and FastNLO for D0
Tevatron Run II measurement \cite{:2008hua}.}

\end{figure}

\begin{figure}[h!]
\begin{centering}
\includegraphics[width=1\textwidth]{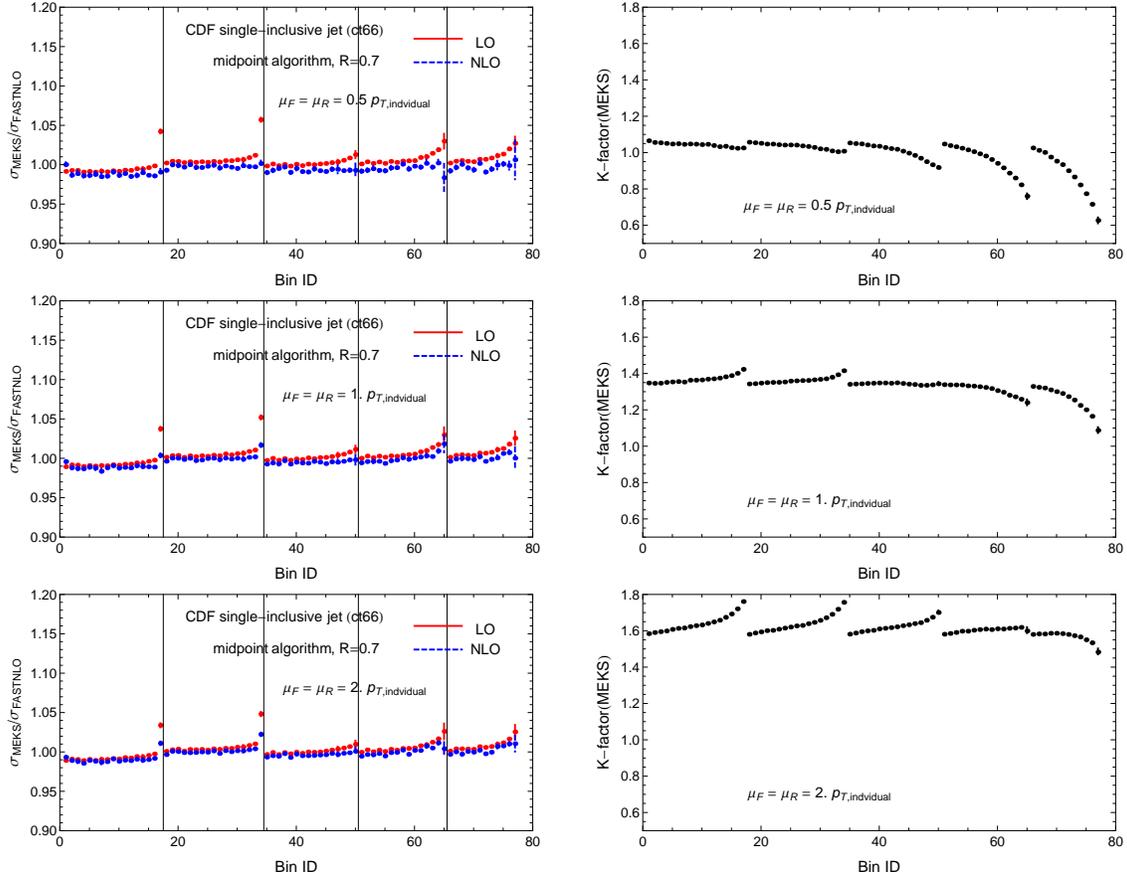}
\par\end{centering}

\caption{\label{jetcom_cdfinc} Comparison of $p_{T}$ distributions
for single-inclusive jet production from MEKS and FastNLO for CDF
Tevatron Run II measurement \cite{Aaltonen:2008eq}.}

\end{figure}

\begin{figure}[h!]
\begin{centering}
\includegraphics[width=1\textwidth]{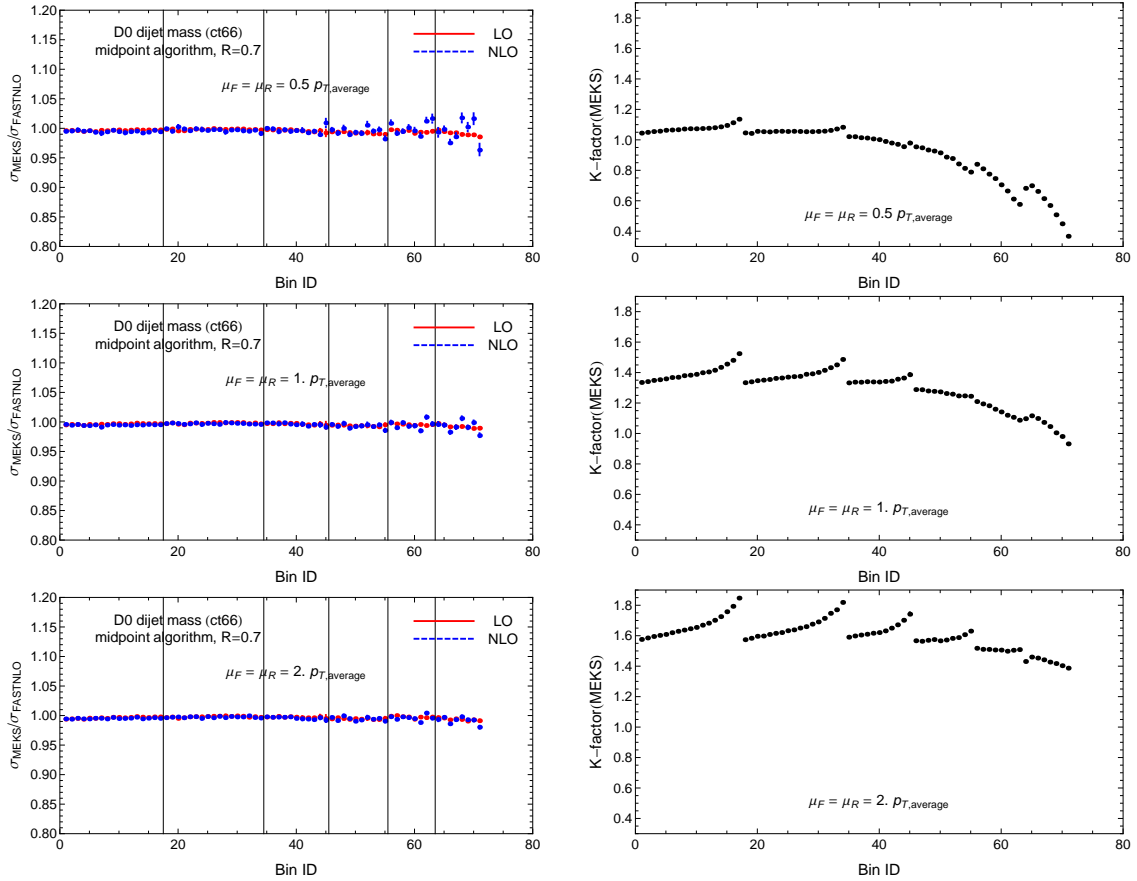}
\par\end{centering}

\caption{\label{jetcom_d0dim} Comparison of invariant mass
distributions for dijet production from MEKS and FastNLO for D0
Tevatron Run II measurement \cite{Abazov:2010fr}.}

\end{figure}

\begin{figure}[h!]
\begin{centering}
\includegraphics[width=1\textwidth]{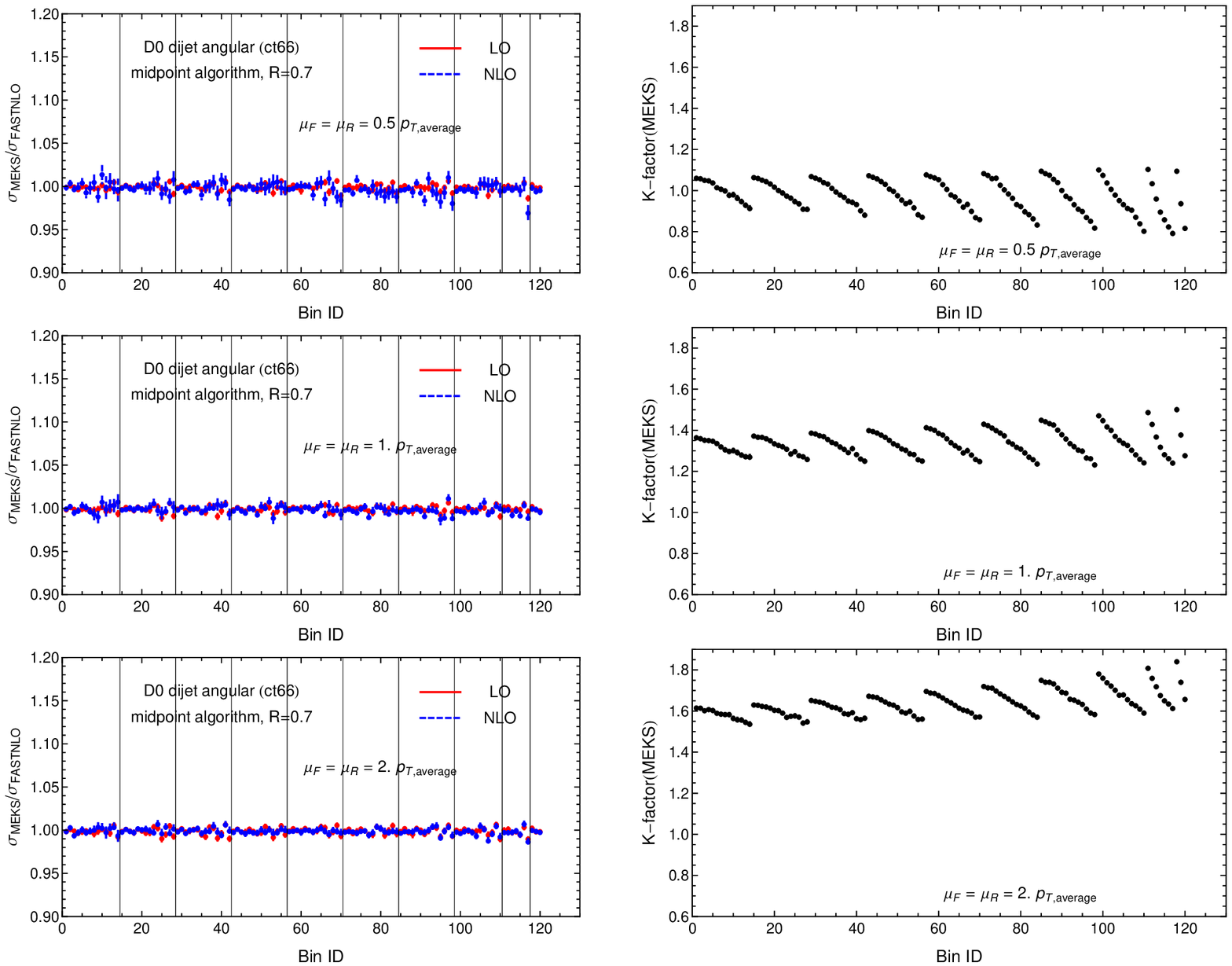}
\par\end{centering}

\caption{\label{jetcom_d0dic} Comparison of angular ($\chi$)
distributions for dijet production from MEKS and FastNLO for D0
Tevatron Run II measurement \cite{:2009mh}.}

\end{figure}

\begin{figure}[h!]
\begin{centering}
\includegraphics[width=1\textwidth]{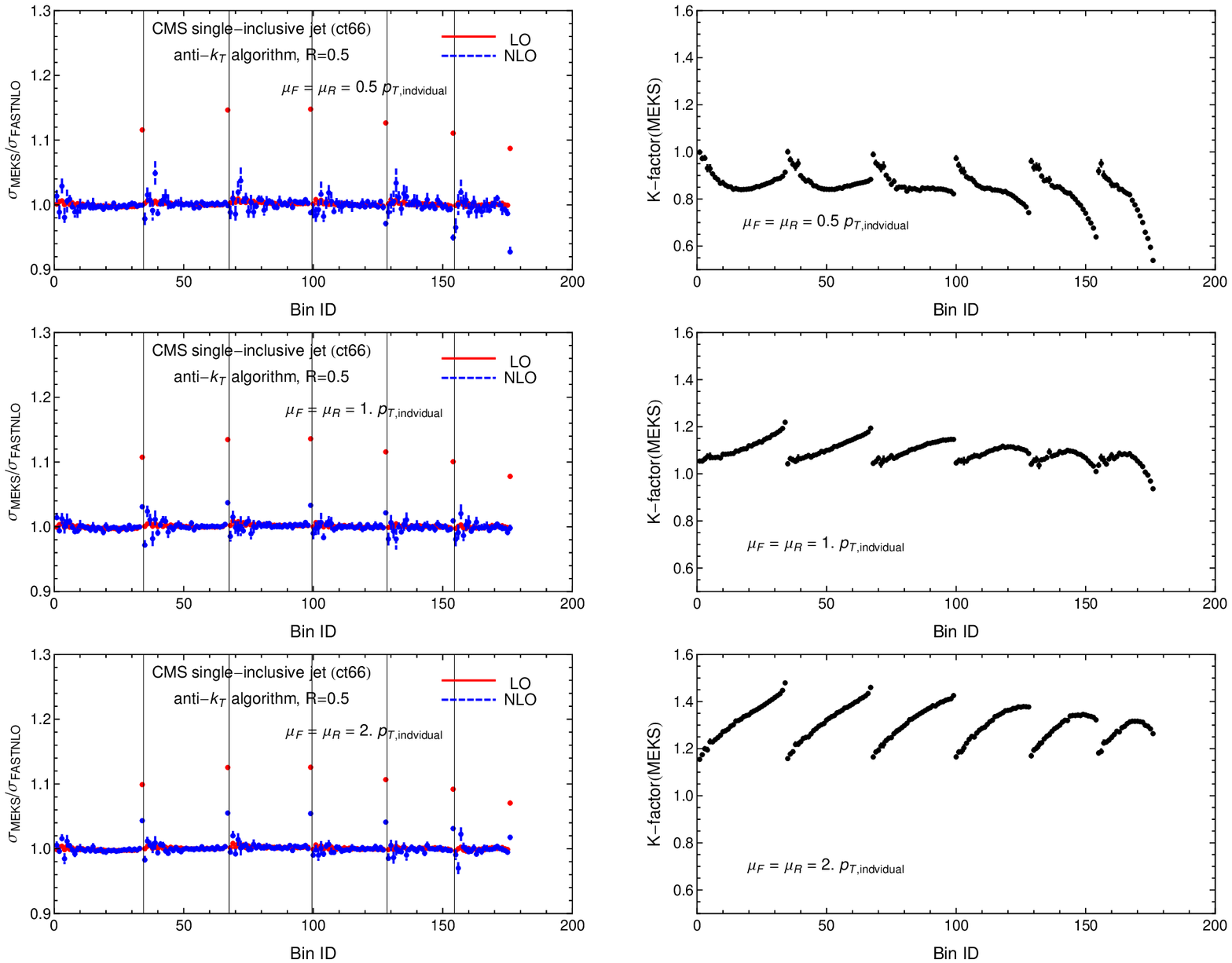}
\par\end{centering}

\caption{\label{jetcom_cmsinc} Comparison of $p_{T}$ distributions
for single-inclusive jet production from MEKS and FastNLO for CMS
LHC (7\,TeV) measurement \cite{CMS:2011ab}.}

\end{figure}

\begin{figure}[h!]
\begin{centering}
\includegraphics[width=1\textwidth]{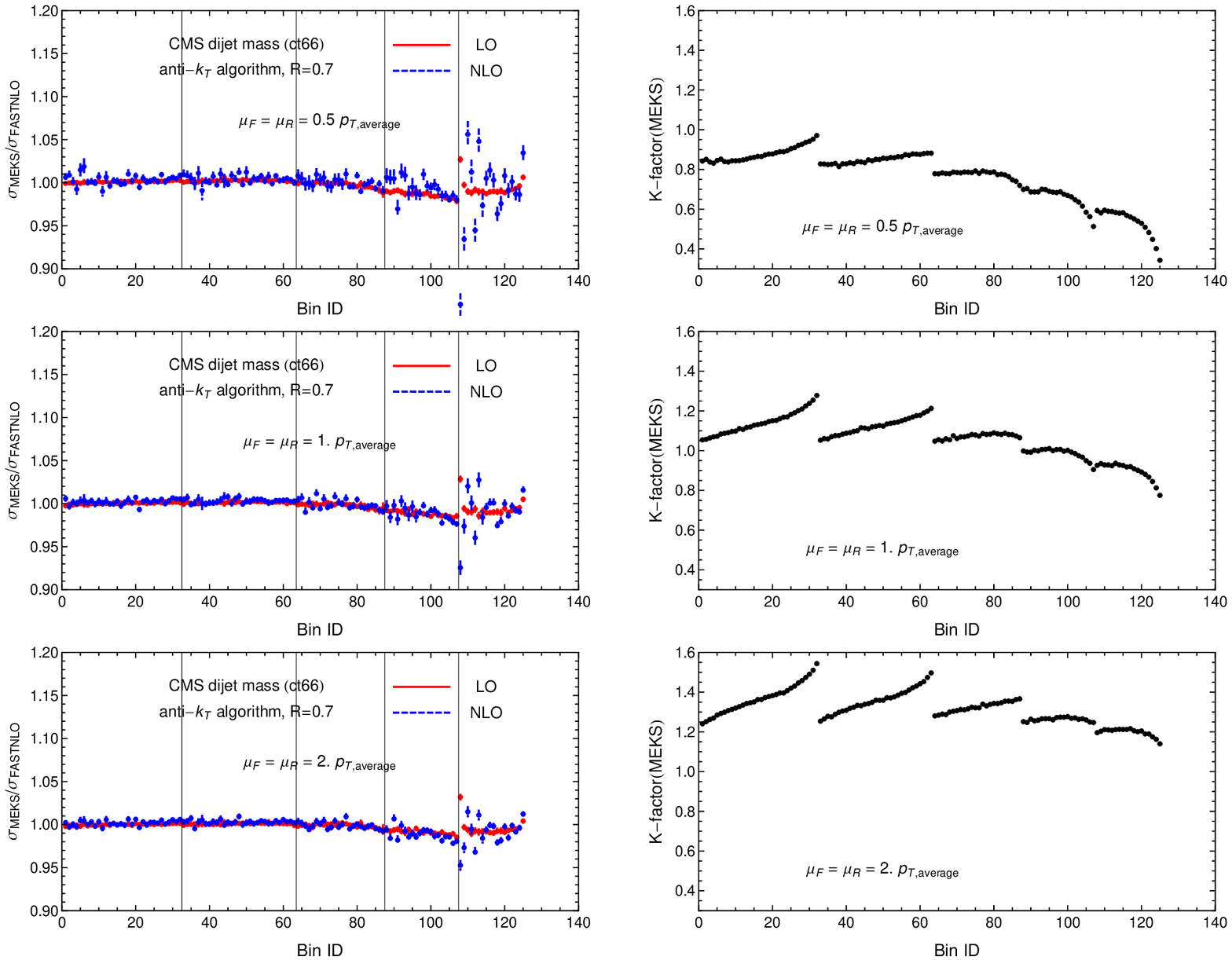}
\par\end{centering}

\caption{\label{jetcom_cmsdim} Comparison of invariant mass
distributions for dijet production from MEKS and FastNLO for CMS LHC
(7\,TeV) measurement \cite{Chatrchyan:2011qta}.}

\end{figure}

Figs.~\ref{jetcom_d0inc}-\ref{jetcom_cmsdim} compare our
representative numerical results with the ones provided by FastNLO
1.0~\cite{Kluge:2006xs} for the $p_{T}$ distributions of
single-inclusive jets, the invariant mass distributions of dijets, and
(in the case of D0 Run-2) the angular distributions ($\chi$) of dijets.
Kinematical bins of the Tevatron ($\sqrt{s}=1.96\mbox{
TeV}$)~\cite{:2008hua,Aaltonen:2008eq,Abazov:2010fr,:2009mh} and
LHC ($\sqrt{s}=7\mbox{ TeV}$)~\cite{Chatrchyan:2011qta,CMS:2011ab}
measurements, and CTEQ6.6 central PDFs~\cite{Nadolsky:2008zw} were
used. For this benchmark comparison, we use the Midpoint algorithm
at the Tevatron and anti-$k_{T}$ algorithm at the LHC. The cone size
$R$ is indicated in the figures. The central $\mu_{F,R}$ scales are
set to the individual jet $p_{T}$ in the single-inclusive jet
production and the average $p_{T}$ of the two leading jets in dijet
production, in accordance with the experimental measurements.
We use the 4D recombination scheme. The comparison to
the $E_{T}$ recombination scheme is included at the end of this
section.

In Table.~\ref{per1}, we summarize the performance of the program
for these representative calculations at NLO as well as LO (shown
by values in parentheses). The elapsed CPU time and the achieved numerical
accuracy are shown for a single job with one iteration run on a 2.5-GHz Intel
Xeon processor. $N_{pts}$ is the total number of experimental bins
for each calculation. The integration errors are averaged over all
the bins. As was mentioned, the user can run parallel jobs and combine
the results at the end, in which case the numerical errors go down
proportionally to $1/\sqrt{N_{tot}}$, where $N_{tot}$ is the total
number of iterations in all jobs. %
\begin{table}[h!]
\caption{\label{per1}Performance of the program for different calculations
with one iteration.}

\centering{}\begin{tabular}{c|lccc}
\hline
Exp. ID & Descriptions  & $N_{pts}$ & CPU time (in mins.) & Numerical errors \tabularnewline
\hline
1  & CDF inclusive jet  & 72  & 75(2.5) & $\sim$8(2)\% \tabularnewline
\hline
2  & D0 inclusive jet  & 110  & 114(3.8) & $\sim$6(2)\% \tabularnewline
\hline
3  & D0 dijet mass  & 71  & 146(3.8) & $\sim$5(2)\% \tabularnewline
\hline
4  & D0 dijet angular  & 120  & 197(3.4) & $\sim$5(1)\% \tabularnewline
\hline
5  & CMS inclusive jet  & 176  & 477(15.0) & $\sim$8(1)\% \tabularnewline
\hline
6  & CMS dijet mass  & 125  & 357(10.7) & $\sim$6(1)\% \tabularnewline
\hline
\end{tabular}
\end{table}

In Figs.~\ref{jetcom_d0inc}-\ref{jetcom_cmsdim}, left panels show
ratios of MEKS to FastNLO cross sections, $\sigma_{{\rm
EKS}}/\sigma_{{\rm FastNLO}}$, at LO (red points with error
bars) and NLO = LO + NLO-correction (blue points with error bars), in
kinematical bins provided by the experiments. The horizontal axis
indicates the ID of each bin, which are arranged in the order of
increasing jet rapidity $y_j$ and then the jet's $p_{Tj}$ or $m_{jj}$ for
the single-inclusive jet and dijet mass measurements, or $m_{jj}$
then $\chi$ for the dijet angular measurement. Vertical lines indicate
the boundaries of each rapidity or invariant mass interval. For
example, Fig.~\ref{jetcom_d0inc} shows $\sigma_{{\rm
EKS}}/\sigma_{{\rm FastNLO}}$ in 6 bins of jet rapidity, with bins
1...23 corresponding to the first rapidity bin ($|y|<0.4$), bins
24...45 corresponding to the second rapidity bin ($0.4<|y|<0.8$),
and so on. The left panel includes, from top to bottom, three plots
obtained with the renormalization and factorization scales equal to
1/2, 1, and 2 times the central scale. We can see a good overall
agreement between MEKS and FastNLO both at LO and NLO. The only
significant discrepancies are found in the highest $p_{Tj}$ bins for
both the Tevatron and LHC single-inclusive jet production, which are
due to the difference in the scale choices used in MEKS and FastNLO
1.0. {[}These differences reduce when going to NLO{]}. In the MEKS
single-inclusive jet calculation, we use the actual $p_{T}$ of the
partonic jet filled into the bin as the scale input. FastNLO 1.0
sets the scale according to a fixed $p_{T}$ value in each
experimental bin, which tends to be different from the actual
$p_{T}$ of the jet in the highest $p_{T}$ bins, which have large
widths. The same reason causes a small normalization shift in other
$p_{T}$ bins. For dijet production, we only observe random
fluctuations at highest $m_{jj}$ that are mainly due to numerical
integration errors. There is a new version of FastNLO, FastNLO
2.0~\cite{Wobisch:2011ij}, in which the scale is no longer set to a
fixed value in each bin. The tables for this version are not
available for the Tevatron jet cross sections, but they are
available for the ATLAS and CMS single-inclusive jet cross sections.
When we used FastNLO 2.0 for the cross sections in
Fig.~\ref{jetcom_cmsinc}, we find much better agreement with our
MEKS results.

As a practical application, we wish to examine the sensitivity of
the NLO cross section to the choice of the QCD scales $\mu_F$ and
$\mu_R$. In the right panels of
Figs.~\ref{jetcom_d0inc}-\ref{jetcom_cmsdim}, for each distribution,
we present plots using MEKS of
the NLO K factor, defined as the
ratio of the NLO differential cross section to the LO one. The value
of the K factor and its stability with respect to the scale choice
may provide an indication of the magnitude of yet higher-order
corrections. To minimize the potential effect of higher-order terms,
one might opt to choose the renormalization and factorization scales
that bring the K factor close to unity in most of the kinematical
region. An alternative approach for setting the scale is based on
the minimal sensitivity method, which suggests to choose the
$\mu_{R}$ and $\mu_{F}$ values (taken to be equal and designated as
$\mu$ in the following) at the point where the scale dependence of
the NLO cross section is the smallest.

In (di)jet production at central rapidities at the Tevatron, both
requirements ($K\approx1$\linebreak and $d\sigma_{NLO}(\mu)/d\mu\approx0$)
could be satisfied by choosing $\mu\approx0.5\, p_{T}$; see, \textit{e.g.},
the appendix in Ref.~\cite{Stump:2003yu}. However, the point of
the minimal sensitivity shifts to higher values (close to $p_{T}$
or even higher) at forward rapidities at the Tevatron or at all rapidities
at the LHC. For such higher scales, however, it is hard to satisfy
the requirement that $K$ remains close to unity at the same time.
This point is illustrated by our plots of the $K$ factors. At the
central rapidities and $\mu_{R}=\mu_{F}=0.5\, p_{T}$ at the Tevatron
(the lowest 3 rapidity bins in Figs.~\ref{jetcom_d0inc}-\ref{jetcom_d0dic}),
$K\approx1$ and is relatively independent of $p_{T}$, as seen in
the top subpanels. However, with this scale choice the $K$ factor
deviates significantly from unity and has strong kinematic dependence
if the rapidity and $p_{T}$ are large. If one chooses the scale that
is equal to $p_{T}$ or even $2\, p_{T}$ (the middle and bottom figures),
in accord with the minimal sensitivity method for the forward bins,
the kinematical dependence of the $K$ factor reduces, but its value
increases to 1.3-1.6 in most of the bins.

For CMS kinematics (Figs.~\ref{jetcom_cmsinc}-\ref{jetcom_cmsdim}),
the $K$ factor has significant kinematical dependence for all central
scale choices, however, the choice $\mu_{R}=\mu_{F}=p_{T}$ (the middle
subpanels) results in a comparatively flatter $K$ factor that is
also closer to unity. We can see that it is hard to find a fixed scale
(or a scale of the type $p_{T}\times(\mbox{a function of\quad}y)$
\cite{Ellis:1992en}) that would simultaneously reduce the magnitude
of the NLO correction and stabilize its scale dependence and kinematical
dependence. The scale $0.5\, p_{T}$ may be slightly more optimal
at the Tevatron, and the scale $p_{T}$ may be slightly better at
the LHC.

\begin{figure}[h!]
\begin{centering}
\includegraphics[width=0.8\textwidth]{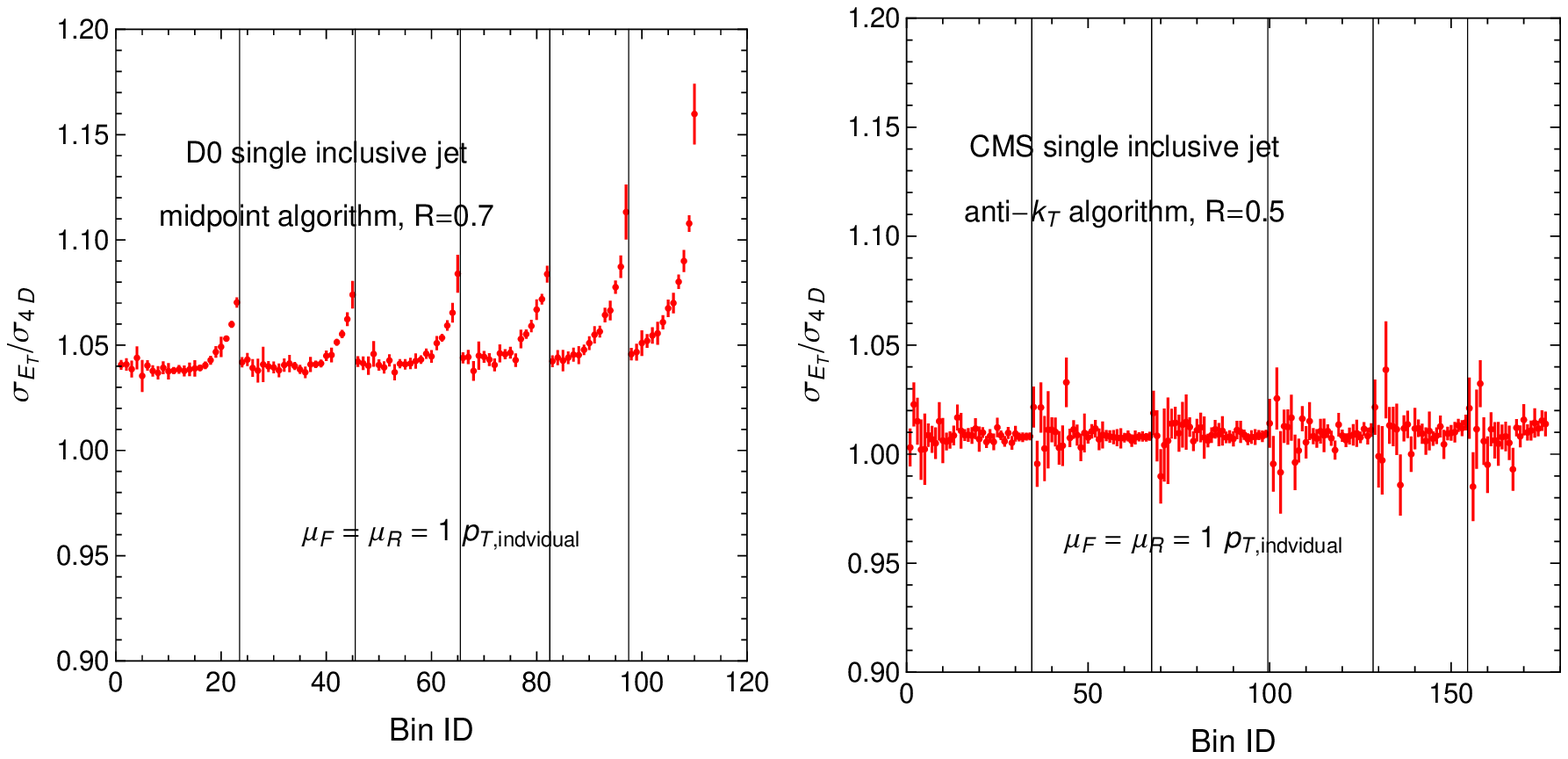}
\par\end{centering}

\caption{\label{jetcom_sinc} Comparison of $p_{T}$ distributions for
single-inclusive jet production using different recombination
schemes.}

\end{figure}

\begin{figure}[h!]
\begin{centering}
\includegraphics[width=0.8\textwidth]{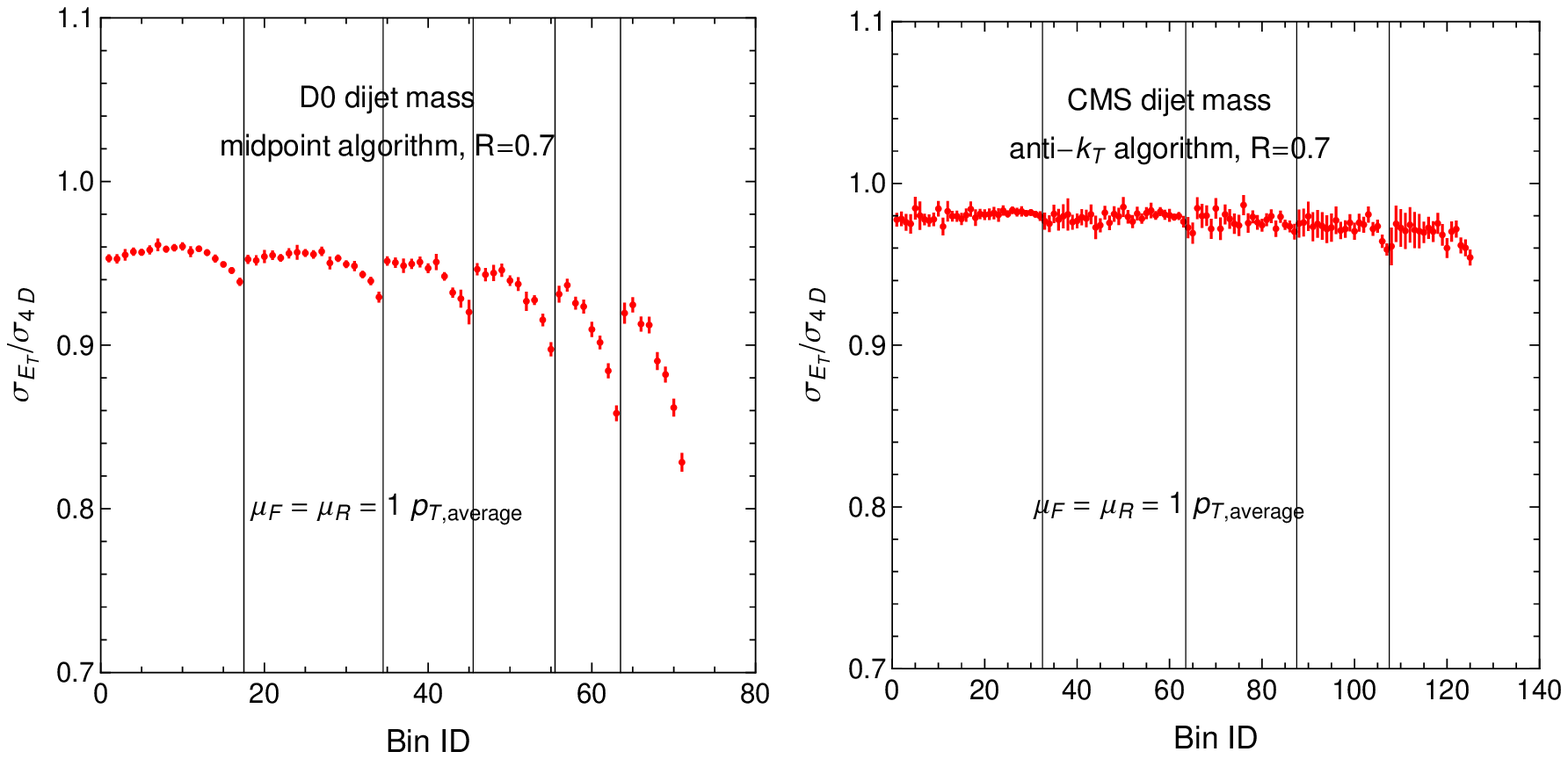}
\par\end{centering}

\caption{\label{jetcom_sdim} Comparison of invariant mass distributions for
the dijet production using different recombination schemes.}

\end{figure}

As a final comparison, in Figs.~\ref{jetcom_sinc} and
\ref{jetcom_sdim}, we plot the ratios of the NLO distributions
calculated using different recombination schemes, where
$\sigma_{4D}$ is obtained with the 4D scheme, and $\sigma_{E_{T}}$
is with the $E_{T}$ scheme. For single-inclusive jet production at
both the Tevatron and LHC, $\sigma_{E_{T}}$ is larger than
$\sigma_{4D}$. An opposite trend is observed in dijet production.
Differences of the predictions based on the two schemes are larger
with the Midpoint algorithm (used at the Tevatron) than with the
anti-$k_{T}$ algorithm (used at the LHC). In a NLO calculation, the
Midpoint algorithm allows a larger maximal angular separation (2$R$)
between the two partons forming a jet, compared to the anti-$k_{T}$
algorithm that only allows the angular separation up to $R$. This
produces the observed different behaviors of the two jet algorithms.

\section{Conclusion \label{sec:Conclusion}}

In conclusion, this document describes the upgraded EKS program
(MEKS) that provides a fast and stable NLO calculation of
double-differential cross sections for single-inclusive jet and
dijet production at hadron colliders. The new program uses the VEGAS
Monte Carlo sampling and the EKS function to generate weighted
events and fill them into finely binned two-dimensional histograms
for a later analysis. It also includes a user interface to add new
jet observables, which is advantageous compared to the popular
FastNLO code \cite{Kluge:2006xs,ftnlo:2010xy} that provides only
the cross sections in the bins of the already completed
measurements. Distributions of sample events are tuned automatically
to speed up convergence of the integration of steep differential
distributions. The program allows parallelization of the Monte-Carlo
integration. In order to facilitate the precision comparison of the
MEKS code with other existing codes for computations of NLO cross
sections, we document a benchmark comparison of the MEKS and
FastNLO, and find good agreement. The MEKS code is available at the
webpage http://www.hepforge.org/downloads/meks.

\subsubsection*{ACKNOWLEDGMENTS}

This work was supported by the U.S. DOE Early Career Research Award
DE-SC0003870 and by Lightner-Sams Foundation; by the U.S. Department
of Energy under Grant No. DE-FG02-96ER40969; by the U.S. National
Science Foundation under Grant No. PHY-0855561; by the National
Science Council of Taiwan under Grant Nos. NSC-98-2112-M-133-002-MY3
and NSC-101-2112-M-133-001-MY3. PMN appreciates helpful discussions
of benchmark comparison with J. Huston, J. Rojo, and M. Wobisch. PMN
and DES also benefited from discussing related work with participants of the
Workshop {}``High Energy QCD at the start of the LHC'' at the
Galileo Galilei Institute of Theoretical Physics (INFN Florence,
September 2011); they thank the organizers of this workshop for
financial support and hospitality. HLL thanks the hospitality of the
department of Physics and Astronomy at Michigan State University,
where part of this work was done.



\providecommand{\href}[2]{#2}\begingroup\raggedright

\endgroup

\begin{thebibliography}{37}
\bibitem{Abazov:2009nc} V.~M. Abazov {\em et.~al.},, \textbf{D0}
Collaboration {\em Phys. Rev.} \textbf{D80} (2009) 111107.

\bibitem{Lai:2010vv} H.-L. Lai {\em et.~al.}, {\em Phys. Rev.}
\textbf{D82} (2010) 074024.

\bibitem{Guzzi:2011sv} M.~Guzzi {\em et.~al.}, \href{http://arXiv.org/abs/arXiv:1101.0561}{\texttt{arXiv:1101.0561
{[}hep-ph{]}}}.

\bibitem{Martin:2009iq} A.~D. Martin, W.~J. Stirling, R.~S. Thorne,
and G.~Watt, {\em Eur. Phys. J.} \textbf{C63} (2009) 189.

\bibitem{Ball:2010de} R.~D. Ball {\em et.~al.}, {\em Nucl.
Phys.} \textbf{B838} (2010) 136--206.

\bibitem{JetCorrelations2012} Z.~Liang and P.~Nadolsky, \href{http://arXiv.org/abs/arXiv:1203.6803}{\texttt{arXiv:1203.6803
{[}hep-ph{]}}}, p. 36.

\bibitem{Chatrchyan:2011ns} S.~Chatrchyan {\em et.~al.},, \textbf{CMS}
Collaboration {\em Phys. Lett.} \textbf{B704} (2011) 123.

\bibitem{Khachatryan:2011as} V.~Khachatryan {\em et.~al.},,
\textbf{CMS} Collaboration {\em Phys. Rev. Lett.} \textbf{106}
(2011) 201804.

\bibitem{Aad:2011aj} G.~Aad {\em et.~al.},, \textbf{ATLAS} Collaboration
{\em New J. Phys.} \textbf{13} (2011) 053044.

\bibitem{:2010wv} G.~Aad {\em et.~al.},, \textbf{ATLAS} Collaboration
{\em Eur. Phys. J.} \textbf{C71} (2011) 1512.

\bibitem{Chatrchyan:2011qta} S.~Chatrchyan {\em et.~al.},, \textbf{CMS}
Collaboration {\em Phys. Lett.} \textbf{B700} (2011) 187.

\bibitem{CMS:2011ab} S.~Chatrchyan {\em et.~al.},, \textbf{CMS}
Collaboration {\em Phys. Rev. Lett.} \textbf{107} (2011) 132001.

\bibitem{Ellis:1992en} S.~D. Ellis, Z.~Kunszt, and D.~E. Soper,
{\em Phys. Rev. Lett.} \textbf{69} (1992) 1496.

\bibitem{Kunszt:1992tn} Z.~Kunszt, and D.~E. Soper, {\em Phys.
Rev.} \textbf{D46} (1992) 192.

\bibitem{Nagy:2001fj} Z.~Nagy, {\em Phys. Rev. Lett.} \textbf{88}
(2002) 122003.

\bibitem{Nagy:2003tz} Z.~Nagy, {\em Phys. Rev.} \textbf{D68}
(2003) 094002.

\bibitem{Kidonakis:2000gi} N.~Kidonakis and J.~F. Owens, {\em
Phys. Rev.} \textbf{D63} (2001) 054019.

\bibitem{Kluge:2006xs} T.~Kluge, K.~Rabbertz, and M.~Wobisch,
\href{http://arXiv.org/abs/hep-ph/0609285}{\texttt{hep-ph/0609285}}.

\bibitem{ftnlo:2010xy} See http://projects.hepforge.org/fastnlo/form/index.html.

\bibitem{Alioli:2010xa} S.~Alioli, K.~Hamilton, P.~Nason, C.~Oleari,
and E.~Re, {\em JHEP} \textbf{04} (2011) 081.

\bibitem{Hahn:2004fe} T.~Hahn, {\em Comput. Phys. Commun.} \textbf{168}
(2005) 78.

\bibitem{Lepage:1977sw}G.~P.~Lepage, \emph{J.~Comput.~Phys.}~\textbf{27}
(1978) 192.

\bibitem{Blazey:2000qt} G.~C. Blazey {\em et.~al.}, \href{http://arXiv.org/abs/hep-ex/0005012}{\texttt{hep-ex/0005012}}.

\bibitem{Cacciari:2008gp} M.~Cacciari, G.~P. Salam, and G.~Soyez,
{\em JHEP} \textbf{04} (2008) 063.

\bibitem{Catani:1993hr} S.~Catani, Y.~L. Dokshitzer, M.~H. Seymour,
and B.~R. Webber, {\em Nucl. Phys.} \textbf{B406} (1993) 187.

\bibitem{Ellis:1993tq}
  S.~D.~Ellis and D.~E.~Soper, {\em Phys. Rev.} \textbf{D48} (1993)
  3160.


\bibitem{Dokshitzer:1997in}
  Y.~L.~Dokshitzer, G.~D.~Leder, S.~Moretti and B.~R.~Webber,
   {\em JHEP} \textbf{08} (1997) 001.

\bibitem{Salam:2009jx} G.~P. Salam, {\em Eur. Phys. J.} \textbf{C67}
(2010) 637.


\bibitem{LHAPDF}M.~R.~Whalley, D.~Bourilkov, and R.~C.~Group,
hep-ph/0508110. The latest LHAPDF library can be downloaded from
http://hepforge.cedar.ac.uk/lhapdf/ .


\bibitem{:2008hua} V.~M. Abazov {\em et.~al.},, \textbf{D0}
Collaboration {\em Phys. Rev. Lett.} \textbf{101} (2008) 062001.

\bibitem{Aaltonen:2008eq} T.~Aaltonen {\em et.~al.},, \textbf{CDF}
Collaboration {\em Phys. Rev.} \textbf{D78} (2008) 052006.

\bibitem{Abazov:2010fr} V.~M. Abazov {\em et.~al.},, \textbf{D0}
Collaboration {\em Phys. Lett.} \textbf{B693} (2010) 531.

\bibitem{:2009mh} V.~Abazov {\em et.~al.},, \textbf{D0} Collaboration
{\em Phys.Rev.Lett.} \textbf{103} (2009) 191803.


\bibitem{Nadolsky:2008zw} P.~M. Nadolsky {\em et.~al.}, {\em
Phys. Rev.} \textbf{D78} (2008) 013004.


\bibitem{Wobisch:2011ij}
  M.~Wobisch {\it et al.}  [fastNLO Collaboration],
  arXiv:1109.1310 [hep-ph].


\bibitem{Stump:2003yu} D.~Stump, J.~Huston, J.~Pumplin, W.-K.
Tung, H.-L. Lai, S.~Kuhlmann, and J.~F. Owens, {\em JHEP} \textbf{0310}
(2003) 046.


\end{thebibliography}
\end{document}